\documentclass[10pt,a4paper]{article}
\usepackage{graphicx}
\usepackage{subfigure}
\usepackage{amsmath,amsthm}
\usepackage{amssymb,amsfonts}
\usepackage{authblk}
\usepackage{url}
\usepackage{hyperref}
\usepackage{booktabs}
\usepackage{makecell}

\newtheorem{definition}{\textbf{Definition}}

\title{Change Point Detection with Copula Entropy based Two-Sample Test}
\author{Jian MA\thanks{Email: majian@hitachi.cn}}
\affil{Hitachi China Research Laboratory}
\date{}

\begin{document}

\maketitle

\begin{abstract}
	\noindent
	Change point detection is a typical task that aim to find changes in time series and can be tackled with two-sample test. Copula Entropy is a mathematical concept for measuring statistical independence and a two-sample test based on it was introduced recently. In this paper we propose a nonparametric multivariate method for multiple change point detection with the copula entropy-based two-sample test. The single change point detection is first proposed as a group of two-sample tests on every points of time series data and the change point is considered as with the maximum of the test statistics. The multiple change point detection is then proposed by combining the single change point detection method with binary segmentation strategy. We verified the effectiveness of our method and compared it with the other similar methods on the simulated univariate and multivariate data and the Nile data.
\end{abstract}

\section{Introduction}
%change point detection
Change point detection is a typical task that aim to find single or multple changes in time series. The detection can be offline or online and the time series can be univariate or multivariate. In this paper, we focus on offline multivariate multiple change point detection. Many algorithms have been proposed for the task, see \cite{Truong2020,Aminikhanghahi2017,Reeves2007,Niu2016} for the reviews on this topic. Change point detection can be widely applied to natural, social, or industrial systems where abrupt changes happen.

%two-sample test
Two-sample test is a common problem of hypothesis testing in statistics. It is to test the hypothesis whether two samples are from a same distribution. There are many two-sample test based on different mathematical concepts. A typical way of defining test statistic for testing is based on the measures of statistical independence in two samples, such as kernels base measure \cite{Gretton2012}, mutual information \cite{Guha2014}.

%copula entropy and tst
Copula Entropy (CE) is a recently defined mathematical concept for measuring statistical independence \cite{Ma2011}. It is proved to be equivalent to mutual information in information theory. A nonparametric method for estimating it was also proposed \cite{Ma2011}. Recently, CE has been applied to two-sample test \cite{Ma2023}, in which the test statistic is defined as the difference between CEs of two hypotheses.

% related work
There are several work on change point detection with copulas. Xiong and Cribben \cite{Xiong2023b} proposed a method for estimating change points with Vine copula and applied it to fMRI data. B\"ucher et al \cite{Buecher2014a} proposed a change point detection method based on empirical copula process. Stark and Otto \cite{Stark2020} proposed to test structural changes in multivariate time series in copula-base dependence measures, such as Spearman's $\rho$ and quantile dependencies.

%this paper
In this paper, we propose to use CE-based two-sample test for multiple change point detection. The idea is simple: first transforming the change point detection problem into a group of CE-based two-sample tests on every points of time series and then find the change point as that with the maximum of the test statistics. A multiple change point detection problem can be solved by combining the single change point detection method with binary segmentation strategy. Since the CE-based two-sample test is nonparametric and multivariate, the proposed change point detection method is also nonparametric and multivariate. We verified the effectiveness of the proposed method and compared it with the other similar methods on both simulated and real data in this paper.

This paper is organized as follows: Section \ref{sec:methodology} introduces copula entropy and the two-sample test based on it, Section \ref{sec:method} presents the proposed methods on single and multiple change point detection, experiments with simulated and real data will be presented in Section \ref{sec:sim} and Section \ref{sec:real} respectively, followed by some discussion in Section \ref{sec:discussion}, and finally we conclude the paper in Section \ref{sec:con}.

\section{Methodology}
\label{sec:methodology}

\subsection{Copula Entropy}
\label{sec:ce}
Copula theory is a probabilistic theory on representation of multivariate dependence \cite{nelsen2007,joe2014}. According to Sklar's theorem \cite{sklar1959}, any multivariate density function can be represented as a product of its marginals and copula density function (cdf) which represents dependence structure among random variables. 

With copula theory, Ma and Sun \cite{Ma2011} defined a new mathematical concept, named Copula Entropy, as follows:
\begin{definition}[Copula Entropy]
	Let $\mathbf{X}$ be random variables with marginals $\mathbf{u}$ and copula density function $c$. The CE of $\mathbf{X}$ is defined as
	\begin{equation}
	H_c(\mathbf{x})=-\int_{\mathbf{u}}{c(\mathbf{u})\log c(\mathbf{u})d\mathbf{u}}.
	\label{eq:ce}
	\end{equation}	
\end{definition}

A non-parametric estimator of CE was also proposed in \cite{Ma2011}, which composed of two simple steps:
\begin{enumerate}
	\item estimating empirical copula density function;
	\item estimating the entropy of the estimated empirical copula density.
\end{enumerate}
The empirical copula density in the first step can be easily derived with rank statistic. With the estimated empirical copula density, the second step is essentially a problem of entropy estimation, which can be tackled with the KSG estimation method \cite{Kraskov2004}. In this way, a non-parametric method for estimating CE was proposed in \cite{Ma2011}.

\subsection{Two-sample test with CE}
\label{sec:test}
CE has been applied to solve the two-sample test problem \cite{Ma2023}. Given two samples $\mathbf{X}_1 =\{X_{11},\cdots,X_{1m}\} \sim  P_1, \mathbf{X}_2 =\{X_{21},\cdots,X_{2n}\}\sim P_2$, the null hypothesis for two sample test is
\begin{equation}
	H_0: P_1 = P_2,
\end{equation}
and the alternative is
\begin{equation}
H_1: P_1 \neq P_2.
\end{equation}
where $\mathbf{X}_1,\mathbf{X}_2 \in R^d$ and $P_1,P_2$ are the corresponding probability distribution functions.

Let $\mathbf{X} = (\mathbf{X}_1, \mathbf{X}_2)$ and $Y_0,Y_1$ be two labeling variables for the two hypotheses respectively that $Y_1=(0_1,\cdots,0_m, 1_1, \cdots, 1_n)$ and $Y_0=(1_1,\cdots,1_{m+n})$. So the CE between $\mathbf{X}$ and $Y_i$ can be calculated as
\begin{equation}
	H_c(\mathbf{X};Y_i) = H_c(\mathbf{X},Y_i)-H_c(\mathbf{X}).
\end{equation}
Then the test statistic for $H_0$ is defined as the difference between the CEs of the two hypotheses, as follows:
\begin{equation}
	T_{ce}(X_1,X_2)=H_c(\mathbf{X},Y_0)-H_c(\mathbf{X},Y_1).
	\label{eq:tce}
\end{equation}
It is easy to know $T_{ce}$ will be a small value if $H_0$ is true and a large value if $H_1$ is true. 

The test statistic in \eqref{eq:tce} can be easily estimated from data by estimating the two terms in it with the non-parametric estimator of CE. Since the CE estimator is non-parametric, the estimator of the test statistic can be applied to any cases without assumptions. Another merit of such estimator of the test statistic is hyperparameter-free.

\section{Proposed Method}
\label{sec:method}

\subsection{Single change point detection}
In this section, we first propose a method for single change point detection based on the above two-sample test. The idea is simple: for a time series, the CE-based two-sample test is conducted on two sub-series devided by each point of time series and the point associated with the maximal test statistic of these tests is the change point.

Given a time series $\mathbf{X}=\{x_1,\ldots,x_n\}, x_i\in R^d$, the single change point detection problem can be formulated as follows:
\begin{equation}
	i = {\underset{i\in [1,n-1]}{{\arg\max}}\, T_{ce}(X_1,X_2)},
\end{equation}
where $T_{ce}(X_1,X_2)$ is the statistic of the CE-based two-sample test on the samples $X_1=\{x_1,\ldots,x_i\}$ and $X_2=\{x_{i+1},\ldots,x_n\}$. 

\subsection{Multiple change point detection}
The multiple change point detection can be transformed as a group of the above single change point detection problems with binary segmentation strategy. For a time series data, we first detect a change point with the above single change point detection method, and if detected, then the whole time series is separated into two segments before and after the detected change point. Such detection is continued on the two segments such derived till no change point can be detected on all the following such derived segments. 

In the proposed method, a threshold on the test statistic is set for judging whether there is a change point in each segment. A change point is detected if its associated maximal test statistic is larger than the threshold. By means of a threshold, our method can estimate the number of multiple change points automatically.

Our method is based on the two-sample test in Section \ref{sec:test}. Since the CE-based test is nonparametric and multivariate, the proposed method is also nonparametric and multivariate, and can be applied to any cases without assumptions.

\section{Simulations}
\label{sec:sim}
\subsection{Experiments}
We conducted simulation experiments to test the proposed method. In each simulation, a univariate or multivariate time series data with several change points was first simulated and our method was then applied to the simulated data to detect these change points. Each time series was composed of four sub-series generated from four different distributions with length as 50 points, which means there are three change points at $[51,101,151]$.

For univariate time series, all the sub-series were generated with normal distribution with different mean and variance. We simulated three typical cases of change points: different means, different means and variances, and different variances. The parameters of the normal distributions in the three cases are listed in Table \ref{tab:uni-sim}.

\begin{table}
	\centering
	\caption{Parameters (mean $\mu$ and variance $\delta$) of the normal distributions in univariate time series simulations.}
	\label{tab:uni-sim}
	\begin{tabular}{l|c|c|c|c|c|c|c|c}
		\toprule
		&$\mu_1$&$\delta_1$&$\mu_2$&$\delta_2$&$\mu_3$&$\delta_3$&$\mu_4$&$\delta_4$\\
		\midrule
		mean&0&1&5&1&10&1&3&1\\
		mean-var&0&1&5&3&10&1&3&10\\
		var&0&1&0&10&0&5&0&1\\
		\bottomrule
	\end{tabular}
\end{table}

For multivariate time series, the sub-series were generated with bi-variate normal distributions with different mean and covariances first. We simulated three typical cases as well: different means, different means and covariances, and different covariances. We also simulated a group of sub-series with bi-variate normal distributions and bi-variate copula functions. The copula functions here are frank copula ($\theta=0.9$) and normal copula ($\rho=0.3$) both with normal distribution ($\mu=0$ and $\delta=2$) and exponential distribution (rate=0.5) as marginals. The parameters of the simulations in the three cases are list in
Table \ref{tab:multi-sim}.

\begin{table}
	\centering
	\caption{Parameters of the normal distributions or copula functions of the four sub-series in multivariate time series simulations.}
	\label{tab:multi-sim}
	\begin{tabular}{l|c|c|c|c}
		\toprule
		&1&2&3&4\\
		\midrule
		mean&\makecell[c]{$\mu=(0,0)$,\\$cov=0.2$}&\makecell[c]{$\mu=(10,10)$,\\$cov=0.2$}&\makecell[c]{$\mu=(5,5)$,\\$cov=0.2$}&\makecell[c]{$\mu=(1,0)$,\\$cov=0.2$}\\
		\hline
		mean-var&\makecell[c]{$\mu=(0,0)$,\\$cov=0.2$}&\makecell[c]{$\mu=(10,10)$,\\$cov=0.8$}&\makecell[c]{$\mu=(5,5)$,\\$cov=0.1$}&\makecell[c]{$\mu=(1,0)$,\\$cov=0.9$}\\
		\hline
		mean-var&\makecell[c]{$\mu=(0,0)$,\\$cov=0.2$}&\makecell[c]{$\mu=(0,0)$,\\$cov=0.8$}&\makecell[c]{$\mu=(0,0)$,\\$cov=0.1$}&\makecell[c]{$\mu=(0,0)$,\\$cov=0.9$}\\
		\hline
		copula&\makecell[c]{$\mu=(0,0)$,\\$cov=0.2$}&\makecell[c]{frank copula}&\makecell[c]{$\mu=(5,5)$,\\$cov=0.1$}&\makecell[c]{normal copula}\\
		\bottomrule
	\end{tabular}
\end{table}

We compared our method with traditional change point detection methods. In univariate three cases, our method and the three methods for detecting changes in mean, in mean and variance, and in variance were compared respectively. The binary segmentation strategy \cite{Scott1974} were adopted in these three compared methods. In multivariate cases, our method was compared with the kernel change point detection method \cite{Arlot2019}. The penalty parameter of the kernel method was tuned to obtain the best possible results.

In the experiments, the implementation of the CE-based two-sample test in the \textbf{R} package \texttt{copent}\cite{Ma2021} was used. The thresholds for the test statistics is 0.13 in all the experiments, except for the multivariate case with different covariance case the threshold is 0.05. The compared method in univariate cases was those implemented in the \textbf{R} package \texttt{changepoint}\cite{Killick2014}. The kernel method implemented in the \textbf{R} package \texttt{ecp}\cite{James2015} was used. The codes of the experiments are available at \url{https://github.com/majianthu/cpd}.

\subsection{Results}
The simulation results on univariate and multivariate time series data are presented in Table \ref{tab:uni-res} and \ref{tab:multi-res} respectively. For the univariate data, our method detected all the change points in different means, different means and variances, and different variances cases, as the compared method did. For the multivariate data, both our method and the kernel method work well in different means, different means and variances cases. In different variances case, our method detected two right change points (48,102) with additional false positives while the kernel method detected only false positives after hyperparameter tuning. In the copula function case, our method detected one change point (155) while the kernel method cannot detect any change point. There are two false positives in the different variances case of univariate data (9) and in the different means case of multivariate case (18) but both test statistics of these false positives are smaller than those of the right change points.

\begin{table}
	\centering
	\caption{Detected change points in univeriate time series simulations.}
	\label{tab:uni-res}
	\begin{tabular}{l|c|c}
		\toprule
		&Our method&Compared method\\
		\midrule
		mean&52,101,151&50,100,150\\
		mean-var&52,101,151&50,100,150\\
		var&9,50,100,151&50,99,150\\
		\bottomrule
	\end{tabular}
\end{table}

\begin{table}
	\centering
	\caption{Detected change points in multivariate time series simulations.}
	\label{tab:multi-res}
	\begin{tabular}{l|c|c}
		\toprule
		&Our method&Kernel method\\
		\midrule
		mean&51,101,151,18&1,51,101,151,201\\
		mean-var&51,101,151&1,51,101,151,201\\
		var&14,48,102,162,169&1,46,59,80,157,159,201\\
		copula&155&1,201\\
		\bottomrule
	\end{tabular}
\end{table}

\section{Real data}
\label{sec:real}
We verified the effectiveness of the proposed method on the Nile data, a well-known benchmark for change point detection \cite{Cobb1978}, which contains the time series measurement of the annual flow of the river Nile at Aswan from 1871 to 1970 with an apparent decreasing change happened at 1898.

\begin{figure}
	\includegraphics[width=\linewidth]{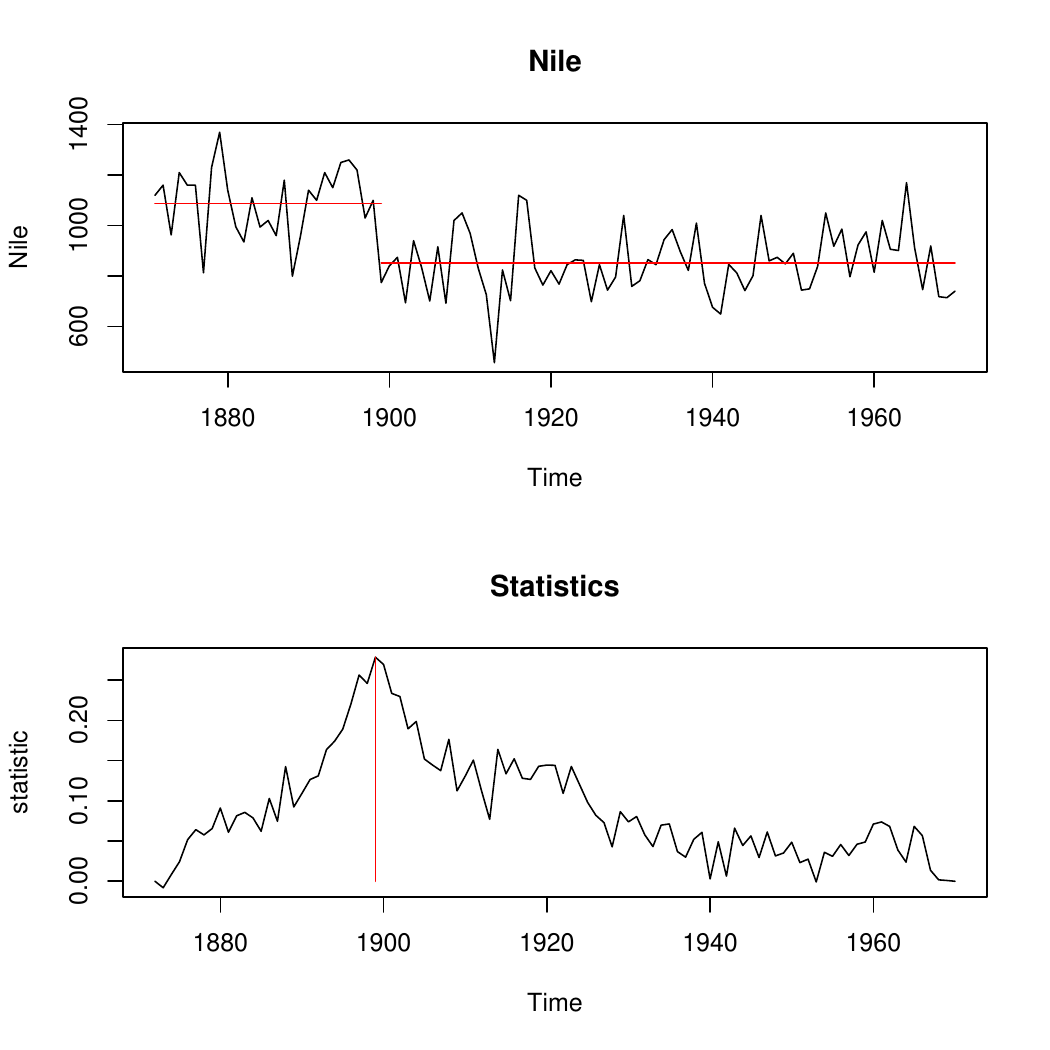}
	\caption{Experimental results on the Nile data.}
	\label{fig:nile}
\end{figure}

We applied the single change point detection method on the Nile data. The results is shown in Figure \ref{fig:nile}, from which it can be learned that our method successfully detect the right point where the change of river flow happened and the test statistic reach it maximum as well.

\section{Discussion}
\label{sec:discussion}
We proposed a method for multiple change point detection with CE-based two-sample test. The power of the proposed method was tested on the simulated and real data. In the simulated experiments, we compared the proposed method with different methods on univariate and multivariate data. Since the proposed method is nonparametric and multivariate, it can be applied directly to all the cases. As contrast, different compared methods should be used for each case of the univariate data.

Our method has one hyperparameter for the threshold on the test statistic. However, in the experiments, only one value (0.13) was used for all the cases, except for the different variances of the multivariate data. It works so well that we did not have to tune it too much. As contrast, the kernel method should tune its penalty parameter frequently for each case to detect the right change points out. This advantage of our method is because that CE is rigorously defined and model-free, and hence the test statistic of the two-sample test based on it is comparable for all the cases.

There are several false positives in the simulation results of our method. However, they can be avoided easily by means of setting larger threshold of the test statistic.

\section{Conclusions}
\label{sec:con}
We propose a nonparametric multivariate method for multiple change point detection with the CE-based two-sample test. The single change point detection is first proposed as a group of two-sample tests on every points of time series data and the change point is considered as with the maximum of the test statistics. The multiple change point detection is then proposed by combining the single change point detection method with binary segmentation strategy. We verified the effectiveness of our method and compared it with the other similar methods on the simulated univariate and multivariate data and the Nile data.

\bibliographystyle{unsrt}
\bibliography{cpd}

\end{document}